\documentclass[aps,prl,onecolumn,showpacs,amsmath,amssymb,floatfixshowpacs, amssymb]{revtex4}
\usepackage{graphicx}
\usepackage{bm}
\usepackage{graphicx}
\usepackage{epsfig}

\def\be{\begin{equation}}
\def\ee{\end{equation}}
\def\bea{\begin{eqnarray}}
\def\eea{\end{eqnarray}}
\def\la{\langle}
\def\ra{\rangle}

\begin{document}
\title{Ultrasensitive Beam Deflection Measurement via Interferometric Weak Value Amplification}
\author{P. Ben Dixon, David J. Starling, Andrew N. Jordan, and John C. Howell}
\affiliation{Department of Physics and Astronomy, University of Rochester, Rochester, New York 14627, USA}
\date{\today}
\begin{abstract}
We report on the use of an interferometric weak value technique to amplify very small transverse deflections of an optical beam.  By entangling the beam's transverse degrees of freedom with the which-path states of a Sagnac interferometer, it is possible to realize an optical amplifier for polarization independent deflections.  The theory for the interferometric weak value amplification method is presented along with the experimental results, which are in good agreement.  Of particular interest, we measured the angular deflection of a mirror down to 560 femtoradians and the linear travel of a piezo actuator down to 20 femtometers.
\end{abstract}
\maketitle

{\it Introduction.}---
It is commonly thought that any average of a quantum operator must be bounded between the smallest and largest of its eigenvalues. This notion was shown to be incorrect by Aharonov, Albert, and Vaidman, who introduce the concept of a {\it weak value} \cite{AAV}.    For example, Aharanov, Albert, and Vaidman described how it would be possible to measure a (post-selected) spin-\( 1/2 \) particle to have \( \la\sigma_z\ra=100 \).  In order to realize this effect, three steps are traditionally carried out: quantum state preparation (pre-selection), a weak perturbation, and post-selection on a final quantum state.  The effect of the the weak perturbation on the post-selected state is measured and usually referred to as the {\it weak measurement}.

The weak value is the average of measurement results over only a sub-set of the data that corresponds to a prescribed outcome of a projective measurement (post-selection).  While initially controversial \cite{review, duck}, the prediction of strange weak values that lay outside the range of the observable's eigenvalues has been experimentally confirmed in the field of quantum optics \cite{Wiseman Photon, First}, and recent proposals exist in condensed matter systems as well \cite{williams-jordan, gefen}.

Weak values are an interesting phenomenon, because they assist us in understanding many counterintuitive quantum results.  For example, weak values can be used as a fundamental test of quantum mechanics by ruling out a class of macrorealistic hidden variable theories and are equivalent to the violation of generalized Leggett-Garg inequalities \cite{williams-jordan}.  Weak values have been useful to help resolve paradoxes that arise in quantum mechanics such as Hardy's paradox \cite{Hardy}, apparent superluminal travel \cite{super}, and more general counterfactual quantum problems such as the three-box problem \cite{review}.

Aside from the fundamental physics interest in weak values, it has been realized that they also are useful.  If we consider the spin of the system as a small signal, the fact that the use of weak values maps this small signal onto a large shift of a measuring device's pointer may be seen as an amplification effect.  Like any amplifier, something must be sacrificed in order to achieve the enhancement of the signal.  For weak values the sacrifice comes in the form of throwing away most of the data in the post-selection process.  
The utility of weak values has been dramatically demonstrated by Hosten and Kwiat \cite{kwiat} who were able to detect a polarization-dependent beam deflection of 1 \AA.  Connections between weak values and other areas of physics include tunable delay lines using fast and slow light in bulk media \cite{FastSlowPhase} and fiber optics \cite{pmd-pdl, fast-light-gisin}, as well as weak value inspired cross phase modulation amplification \cite{Phase Jump}.

This Letter describes the development of a weak value amplification technique for any optical deflection.  In particular, our weak value measurement uses the which-path information of a Sagnac interferometer, and can obtain dramatically enhanced resolution of the deflection of an optical beam.  This technique has several advantages for amplification.  First, it can be used with a variety of beam deflection sources, it is not limited to a polarization dependent deflection.  Second, the post-selection consists simply of a photon emerging from a specified interferometer port.  Finally, the post-selection attenuation is completely independent of the source of deflection because it originates from the destructive interference between the two paths.  In the experiment reported here, the weak measurement consists of monitoring the transverse position of the photon, which gives partial information about the system.

\begin{figure}
\epsfig{file=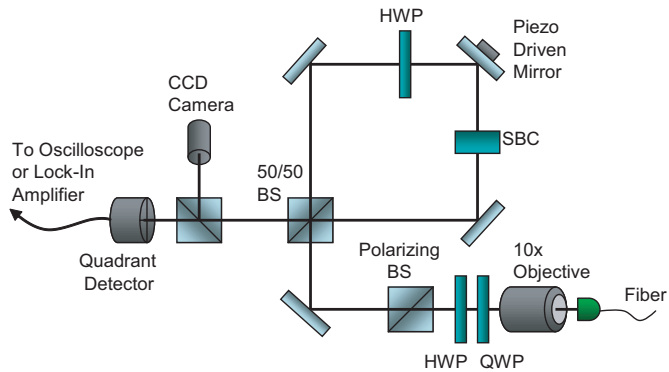, scale=0.7}\caption{Experimental Setup.  The objective lens collimates a 780 nm beam.  After passing through polarization optics, the beam enters a Sagnac interferometer consisting of three mirrors and a 50/50 beamsplitter arranged in a square.  The output port is monitored by both a quadrant detector and a CCD camera.  The SBC and half-wave plate in the interferometer allow the output intensity of the interferometer to be tuned.  The piezo mirror gives a small beam deflection.}
\label{experiment}
\end{figure}

{\it Theoretical Description.}---Consider the schematic of the weak value amplification scheme shown in Fig.~\ref{experiment}.  A light beam enters a Sagnac interferometer composed of a 50/50 beamsplitter and mirrors to cause the beam to take one of two paths and eventually exit the 50/50 beamsplitter.  For an ideal, perfectly aligned Sagnac interferometer, all of the light exits the input port of the interferometer, it is therefore referred to as the bright port, the other port as the dark port.  The constructive interference at the entrance port occurs due to two \( \pi/2 \) phase shifts which occur on reflection in the beamsplitter.  This symmetry is broken with the presence of a half-wave plate and a Soleil-Babinet compensator (SBC), which introduce a relative phase \( \phi \) between the paths, allowing one to continuously change the dark port to a bright port.  While the theory presented is for single photons, the experiment was realized with macroscopic beams.  The effects described here can be understood semi-classically or quantum mechanically, however the amplification effects are identical.

The beam travels through the interferometer, and the spatial shift of the beam exiting the dark-port is monitored.  We refer to the beam's which-path information as the {\it system}, described with the states  \( \{|\!\! \circlearrowright\ra, |\!\! \circlearrowleft \ra\} \).  The transverse position degree of freedom of the beam, labeled by the states \( |x\ra \), is referred to as the {\it meter}. A slight tilt is given to the mirror at the symmetric point in the interferometer.  This tilt corresponds to a shift of the transverse momentum of the beam.  Importantly, the tilt also breaks the symmetry of the Sagnac interferometer, with one propagation direction being deflected to the left of the optical axis at the exit of the beamsplitter, and the other being deflected to the right.  In other words, the which-path observable is coupled to the continuous transverse deflection.

This effect entangles the {\it system} with the {\it meter} via an impulsive interaction Hamiltonian, leading to an evolution operator exp\((-i x {\bf A} k)\) , where \( x \) is the transverse position of the meter, \( k \) is the transverse momentum shift given to the beam by the mirror, and the {\it system} operator \( {\bf A} = |\!\!\circlearrowright\ra \la \circlearrowright\!\!| - |\!\!\circlearrowleft \ra \la \circlearrowleft\!\! | \) describes the fact that this momentum-shift is opposite, depending on the propagation direction.

The splitting of the beam at the 50/50 beamsplitter, plus the SBC (causing the relative phase \( \phi \)) results in an initial {\it system} state of \( |\psi_i\ra = ( i e^{i \phi/2}  |\!\!\circlearrowleft \ra +  e^{-i \phi/2} |\!\!\circlearrowright\ra)/\sqrt{2} \).  The entangling of the position degree of freedom with the which-path degree of freedom results in the state
\be
|\Psi\ra = \int dx \psi(x) |x\ra \exp(-i x {\bf A} k) |\psi_i\ra,
\label{evolution}
\ee
where \( \psi (x) \) is the wavefunction of the {\it meter} in the position basis.  This evolution is part of a standard analysis on quantum measurement, where the above transformation would result in a momentum-space shift of the {\it meter}, \( \Phi(p) \rightarrow \Phi(p \pm k) \), if the initial state is an eigenstate of \( \bf A \).

The weak value analysis then consists of expanding \( \exp(-i x {\bf A} k) \) to first order (assuming \( k a<1 \), where \( a = \sqrt{\la x^2 \ra} \) is the beam initial size) and post-selecting with a final state \( | \psi_f \ra = (|\!\!\circlearrowleft \ra +  i |\!\!\circlearrowright\ra)/\sqrt{2} \) (describing the dark-port of the interferometer).  This leaves the state as
\be
\la \psi_f |\Psi\ra = \int dx \psi(x) |x\ra  [ \la \psi_f |\psi_i\ra - i k x \la \psi_f | {\bf A} |\psi_i\ra].
\label{smallexp}
\ee
We now assume that \( k a | \la \psi_f | A | \psi_i \ra | < | \la \psi_f | \psi_i \ra | < 1 \), and can therefore factor out the dominant state overlap term to find
\be
\la \psi_f |\Psi \ra= \la \psi_f |\psi_i\ra \int dx \psi(x) |x\ra  \exp(-i x A_w k),
\label{reexp}
\ee
where we have re-exponentiated to find an amplification of the momentum shift by the {\it weak value}
\be
A_w = \frac{\la \psi_f | {\bf A} |\psi_i\ra}{\la \psi_f  |\psi_i\ra}
\label{wv}
\ee
with a post-selection probability of \( P_{ps} =|\la \psi_f  |\psi_i\ra|^2  = \sin^2 (\phi/2) \).  The new momentum shift \( k A_w \) will be smaller than the width of the momentum-space wavefunction, \( 1/a \), but the weak value can greatly exceed the \( [-1, 1] \) eigenvalue range of \( {\bf A} \).  In the case at hand, the weak value is purely imaginary, \( A_w = - i \cot (\phi/2) \approx  -  2 i/\phi \) for small \( \phi \).  This has the effect of causing a shift in the position expectation,
\be
\la x \ra = 2 k a^2 |A_w| \approx 4 k  a^2 /\phi,
\label{deflection}
\ee
assuming a symmetric spatial wavefunction.

We can extend this collimated beam analysis and consider putting a lens with a negative image distance \( s_i \) before the interferometer.  This corresponds to a diverging beam.  Using paraxial beam propagation and assuming the initial collimated beam radius \( a \) is significantly larger than the wavelength of light used, the result analogous to Eq.~(\ref{deflection}) is found to have an additional factor of \( F = \ell_{im} (\ell_{im}+\ell_{md})/s_i^2 \), where \( \ell_{im} \) is the distance from the lens image to the moving mirror, and \( \ell_{md} \) is the distance from the moving mirror to the detector \cite{jordan}.

From an experimental point of view, it is convenient to express the deflection in terms of easily measurable quantities.  This can be done with the beam size at the detector \( \sigma = a (\ell_{im}+\ell_{md})/s_i \) and the initial beam size at the lens \( a \) to eliminate \( s_i \) from the equation, and express it in terms of \( \ell_{lm} \), the distance from lens to the moving mirror.  This gives
\be
\la x\ra = 2 k |A_w| \frac{ \sigma^2  \ell_{lm} + \sigma a \ell_{md} }{\ell_{lm}+\ell_{md}}.
\label{exptdeflection}
\ee

Finally, we compare this result to the unamplified deflection (without the interferometer) of \( \delta = k \ell_{md}/k_0 \), where \( k_0 \) is the wavenumber of the light so that \( \theta = k/k_0 \) is the small angle the mirror imparts to the light beam.  The amplification factor is \( {\cal A} = \la x\ra/\delta \).

{\it Experiment.}---
A fiber coupled 780 nm laser beam is collimated using a 10x microscope objective.  Just after the objective, the beam has a Gaussian radius of \(a=\) 640 \(\mu\)m.  The beam can be made to be converging or diverging by moving the fiber end relative to the microscope objective.  After collimation, the beam passes through a polarizing beamsplitter giving a pure horizontal polarization. Half- and quarter-wave plates are used to adjust the intensity of the beam passing through the polarizing beamsplitter. The beam then enters a Sagnac interferometer input port (the pre-selection process). Passing through the interferometer in the clockwise direction, the beam first passes through a half-wave plate which rotates the polarization to vertical, the beam then passes through a SBC which adds a tunable phase to the beam (the SBC is set to add this phase to vertically polarized beams relative to those polarized horizontally). Passing counterclockwise, the beam first passes through the SBC which now has no relative effect, then through the half-wave plate, changing the polarization to vertical.
A piezo electric actuator scans the tilt of one of the interferometer mirrors back and forth.  A gimbal mount is used so that the center of the mirror is the fulcrum.  The tilt of the mirror gives the two propagation directions opposite deflections.  The small beam deflection is the weak interaction between transverse beam deflection ({\it meter}) and which path degree of freedom ({\it system}).

Post-selection is achieved simply by monitoring the light that exits the dark port of the interferometer.  Tuning the SBC to add a small but nonzero relative phase allows a small amount of light out of the dark port. This light is split by a 50/50 beamsplitter and sent to a CCD camera (Newport model LBP-2-USB) which monitors the beam structure, and to a 10 mm by 10 mm  quadrant detector (New Focus model 2921) which monitors beam deflection as well as total power.  The active area of the quadrant detector is significantly larger than the beams used.

The interferometer is roughly square with sides of approximately 15 cm.  The distance from the microscope objective to the piezo driven mirror is \( \ell_{lm} = 48\) cm.  The distance from the piezo driven mirror to the detectors is \( \ell_{md} = 114\) cm (the same distance to both the CCD camera and the quadrant detector). The piezo driven mirror has a lever arm of 3.5 cm.

Piezo deflection was calibrated by removing the 50/50 beamsplitter from the interferometer and observing beam centroid position on the CCD camera.  In this configuration the beam experiences no interference and ray optics describes the beam deflection.  Driving the mirror, the piezo response was found to be 91 pm/mV. The piezo response was verified from 500 Hz down to D.C.

\begin{figure}
\epsfig{file=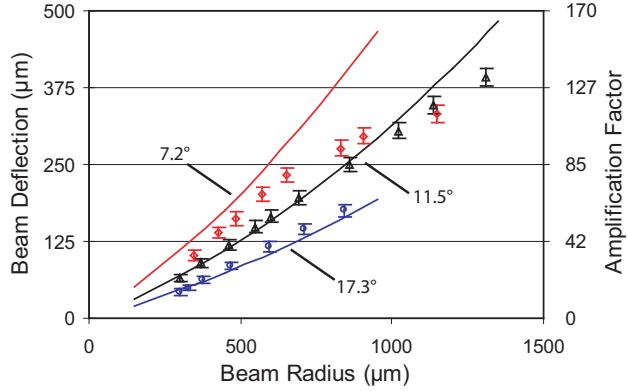, scale=0.93}\caption{Measured beam deflection is plotted as a function of beam radius \(\sigma\).  SBC angle \(\phi\) for each data set is labeled.  The scale on the left is the measured beam deflection \(\la x\ra\).  The scale on the right is the amplification factor \(\cal A\).  The unamplified deflection is \(\delta=2.95\;\mu\)m.  The solid lines are theoretical predictions based on Eq.~(\ref{exptdeflection}).}
\label{characterizationplot}
\end{figure}

To characterize the system the interferometer was first aligned well, minimizing the light exiting the dark port.   The SBC relative phase was then tuned away from zero, allowing light to exit the interferometer.  The piezo driven mirror was given a 500 mV peak to peak amplitude, 100 Hz, sinusoidal driving voltage and the beam deflection was observed using the quadrant detector connected to an oscilloscope.  This was done over a range of beam sizes \(\sigma\) (Gaussian beam radius at the detector), for three values of SBC phase difference.  For these measurements the beam power entering the interferometer was 3.2 mW and the power exiting was between 30 \( \mu \)W and 170 \( \mu \)W.  These measurements, as well as corresponding theoretical prediction curves given by Eq.~(\ref{exptdeflection}) are shown in Fig.~\ref{characterizationplot}.  The measured data is, in general, well described by the theory.

\begin{figure}
\epsfig{file=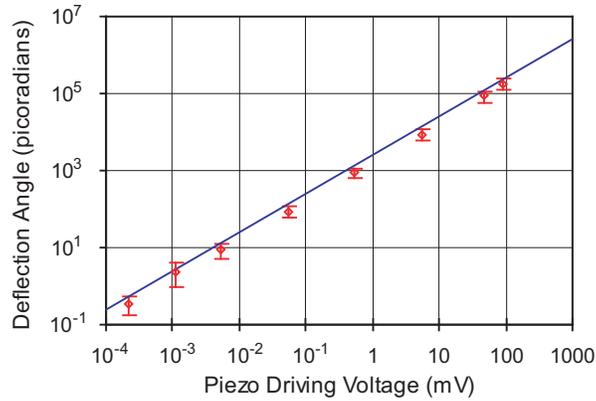}\caption{Angular displacement of the mirror is plotted versus piezo driving voltage.  Weak value signal amplification allows small deflections to be measured.  The solid line shows the expected deflection based on an interpolation of calibrated measurements of the piezo actuator's linear travel at higher voltages.  These data were taken using a weak value amplification of approximately 86.} \label{deflectionplot}
\end{figure}

At the smallest SBC angle (7.2\(^{\circ}\)) the small overlap between pre- and post-selected states allows only a small amount of light to exit the dark port.  With this light at low intensities it begins to be of roughly equal intensity to stray light incident on the quadrant detector.  This leads to less than ideal amplification, as shown in Fig.~\ref{characterizationplot}.  The error bars take into account random error only, not systematic error such as this.

For fixed interferometer output intensity, the range of detectable deflections was also explored.  The interferometer was again aligned such that the beams only had a small phase offset from each other.  For these measurements the beam size at the detector was \(\sigma=1240 \pm 50 \; \mu\)m.  The weak value amplification factor was approximately 86, while the post-selection probability was about 2\% (3.2 mW entered the interferometer while 63 \( \mu \)W exited the dark port).  The amplification factor was found by driving the piezo with a 500 mV peak to peak signal and comparing the measured beam deflection with the aligned interferometer to the measured beam deflection with the interferometer beamsplitter removed.  The piezo driving voltage was varied over five orders of magnitude while the output of the quadrant detector was sent to a lock-in amplifier and the signal was observed.  The smallest driving voltage that yielded measurable a beam deflection was 220 nV corresponding to an angular deflection of the mirror of \(560 \pm 40\) frad (the mirror angle is half the beam deflection angle).  These measurements are shown in Fig.~\ref{deflectionplot}.  At smaller driving voltages, the lock-in amplifier was unable to lock to the signal.

There are other, perhaps more interesting points.   The deflection indirectly measured the linear travel resolution of the piezo electric actuator.  The piezo actuator moved \(20 \pm 2\) fm in making this measurement.  This distance is on the order of large atomic nucleus diameters (Uranium is 15 fm) and is almost six orders of magnitude more resolution than the manufacturer's specifications of 10 nm.  Also, as Hosten and Kwiat point out \cite{kwiat}, weak value measurement techniques such as the one described here reduce technical noise (thermal, electrical, vibrational, etc.).  We are further investigating the topic of reduced technical noise and increased signal to noise ratio.  Further improvements to the system may include: using a quadrant detector with a larger active area which allows a larger beam size to be used, decreasing stray light on the detector by carefully minimizing any back reflections from optics, and aligning the interferometer to have an improved dark port, possibly by using a deformable mirror. As a note, this system may be used for active feedback stabilization since the sinusoidal deflection results in an in-phase sinusoidal amplified signal.

{\it Concluding Remarks.}---
In this paper we have described and demonstrated a method of amplifying small beam deflections using weak values.  The amplification is independent of the source of the deflection.  In this experiment a small mirror deflection in a Sagnac interferometer provides the beam deflection.  By tuning the interferometer and monitoring the resulting small amount of light exiting the interferometer dark port, weak value amplification factors of over 100 are achieved.  The weak value experimental setup, in conjunction with a lock-in amplifier, allows the measurement of 560 frad of mirror deflection which is caused by 20 fm of piezo actuator travel.

This work was supported by DARPA DSO Slow Light, a DOD PECASE award, and the University of Rochester.

\end{document}